\documentclass[12pt]{article}
\usepackage{amsmath}
\usepackage{graphicx}
\usepackage{natbib}
\usepackage{url} 
\usepackage{graphicx}
\usepackage{epsfig}
\usepackage{amsmath}
\usepackage{amssymb}
\usepackage{caption}
\usepackage{color}
\usepackage{enumerate}
\usepackage{multicol}
\usepackage{float}
\usepackage{subfig}
\usepackage{appendix}
\usepackage{setspace}
\usepackage{amsthm}

\newcommand{\blind}{1}

\newcommand{\be}{\begin{equation}}
\newcommand{\ee}{\end{equation}}
\newcommand{\ba}{\begin{eqnarray}}
\newcommand{\ea}{\end{eqnarray}}
\newcommand{\bpm}{\begin{pmatrix}}
	\newcommand{\epm}{\end{pmatrix}}

\newcommand{\beq}{\begin{equation}}
\newcommand{\eeq}{\end{equation}}
\newcommand{\bea}{\begin{eqnarray}}
\newcommand{\eea}{\end{eqnarray}}
\newcommand{\p}{^{\prime}}
\newcommand{\bs}{\boldsymbol}

\newcommand{\nt}{\notag}

\newcommand{\bstheta}{\bs{\theta}}

\usepackage{algorithm}
\usepackage[noend]{algpseudocode}
\usepackage{algorithmicx}
\def\BState{\State\hskip-\ALG@thistlm}

\begin{document}

	\def\spacingset#1{\renewcommand{\baselinestretch}%
		{#1}\small\normalsize} \spacingset{1}

	
	\if1\blind
	{
		\title{\bf A New Framework for Inference on Markov Population Models}
		\author{Adam Walder\\
			Department of Statistics, The Pennsylvania State University, University Park \\
			and \\
			Ephraim M. Hanks \\
			Department of Statistics, The Pennsylvania State University, University Park}
		\maketitle
	} \fi
	
	\if0\blind
	{
		\bigskip
		\bigskip
		\bigskip
		\begin{center}
			{\LARGE\bf A New Framework for Inference on Markov Population Models}
		\end{center}
		\medskip
	} \fi
	
	\bigskip
	\begin{abstract}

		In this work we construct a joint Gaussian likelihood for approximate inference on Markov population models. We demonstrate that Markov population models can be approximated by a system of linear stochastic differential equations with time-varying coefficients. We show that the system of stochastic differential equations converges to a set of ordinary differential equations. We derive our proposed joint Gaussian deterministic limiting approximation (JGDLA) model from the limiting system of ordinary differential equations. The results is a method for inference on Markov population models that relies solely on the solution to a system deterministic equations. We show that our method requires no stochastic infill and exhibits improved predictive power in comparison to the Euler-Maruyama scheme on simulated susceptible-infected-recovered data sets. We use the JGDLA to fit a stochastic susceptible-exposed-infected-recovered system to the Princess Diamond COVID-19 cruise ship data set.   
	
	\end{abstract}

	\noindent%
	{\it Keywords: Stochastic SIR/SEIR;Stochastic Differential Equations; Inference for Mechanistic Models; Epidemiology} 
	\vfill

	\newpage
	\spacingset{1.5} 

	\section{Introduction} 

	Statistical inference methods for population process models are required to understand the dynamics of disease outbreaks such as the current COVID-19 global pandemic \citep{chen2020mathematical,he2020seir,mwalili2020seir}. Markov population models are commonly used to model disease dynamics in small to moderate sized populations \citep{sun2015parameter, allen2017primer, fricks2018stochastic} but simulation and inference can be computationally burdensome. In this work, we construct an approximate joint Gaussian likelihood for Markov population models that helps facilitate inference and simulation. This approximation, which we call the joint Gaussian deterministic limiting approximation (JGDLA) makes inference on Markov population models easier as the joint likelihood of all time-referenced observations can be computed using only the solution to a system of ordinary differential equations (ODEs). This removes the need of for stochastic infill, as is commonly needed for inference on Markov population models.

	A common framework for inference on Markov population models is to consider a Gaussian approximation of the Markov population model using the functional central limit theorem (FCLT). This approximates the Markov population model as a system of linear stochastic differential equations (SDEs). Except in a few specific cases, systems of linear SDEs with time-varying coefficients do not have analytical solutions. In the absence of an analytic solution, numerical methods are required to solve the system of SDEs. The Euler-Maruyama scheme is the most commonly used numerical method for performing inference on SDE models \citep{sun2015parameter,allen2017primer,eisenhauer2020lattice}. The Euler-Maruyama scheme relies on a fixed length time-lag that requires stochastic infill estimates to perform inference or predict at unobserved locations leading to computational bottlenecks.   
	
	The main novel contribution of this work is the construction of the JGDLA model, which is built on the premise that the joint distribution of an It\'o diffusion approximation to the Markov population model can be fully constructed from the solution to a system of ODEs. The result is an approximate inference method for Markov population models which does not require stochastic infill and offers inference familiar to ODE modeling. This is not the first presentation of such a model: \cite{kurtz1978strong,kurtz1981approximation} introduced the mathematical framework needed to justify converge of the approximation, and \cite{baxendale2011sustained} used the approximation technique to investigate the behavior of stochastic population models through simulation. However, previous work only considers simulation and exploration of this approximation. In this work, we develop methods for statistical inference using the JGDLA.    
	
	The population models considered in this work have a direct application in the field of epidemiology \citep{allen2003comparison}. We demonstrate that JGDLA outperforms a Euler-Maruyama scheme on simulated susceptible-infected-recovered (SIR) data sets of varying population sizes in terms of mean absolute prediction error at infill locations. The results of this simulation study show that JGDLA offers inference that only relies on solving an ODE system, does not require stochastic infill for predictions and inference, and provides an improved model fit in comparison to Euler-Maruyama, the most common approximation method as a framework for statistical inference.    
	
	The remainder of the manuscript is organized as follows. In Section \ref{section::Markov Population Models} we introduce Markov population models as the sum of Poisson processes with stochastic rates. In Section \ref{section:Diffusion Approximaitons} we use the FCLT to form a system of linear SDEs for approximating Markov population models. We discuss the Euler-Maruyama scheme for approximate inference on Markov population models in Section \ref{section::Euler-Maruyama Approximation}. In Section \ref{section::Approximation by Limiting Deterministic Sytstem}, we introduce the JGDLA in the statistical framework. We compare JGDLA and the Euler-Maruyama approximation on simulated SIR data sets in Section \ref{section::Simulation Study}. In Section \ref{section::Covid Data Example}, we use the JGDLA to fit a stochastic SEIR model to the Princess Diamond Cruise COVID-19 data set. We conclude with a discussion in Section \ref{section::Discussion}.

	\section{Markov Population Models} \label{section::Markov Population Models}
	
	Deterministic population models are widely used in fields such as chemistry, ecology, and epidemiology to model large scale population dynamics such as disease outbreaks \citep{keeling2011modeling,fricks2018stochastic}. While deterministic models work well for very large populations, stochastic methods are needed to capture fine scale dynamics for populations with few individuals \citep{fricks2018stochastic}. In this section we formulate the Markov population model in terms of stochastic reactions and rates. Our treatment follows the formulation of \cite{kurtz1978strong} and \cite{fricks2018stochastic}. 
	
	Let $\bs{X}(t) = \left(X_1(t),X_2(t),...,X_d(t)\right)\p$ be a d-dimensional random vector on the non-negative integers. $X_j(t)$ represents the number of individuals in a population of size $N$ belonging to subpopulation $j$ at time $t$ (e.g. number of infected individuals in a population). A population reaction occurs when one individual moves from one class to another. There are $n$ possible reactions for any given model, and $d$ subpopulations or classes. We let $\bs{R}_i$ be a d-dimensional vector denoting the $i^{th}$ reaction. For example, if an individual can move from class 1 to 3, the reaction vector will contain $-1$ as the first element, 1 as the third element, and 0 for all other $d-2$ elements.  
	
	Each individual reaction is assumed to occur at a stochastic rate which depends on the current state $\bs{X}(t)$ and unknown rate parameters $\bs{\theta}$. We denote the reaction rate corresponding to $\bs{R}_i$ by $\lambda^{i}_{\bs{\theta}}\left(\bs{X}(t)\right)$. Let $Y_i(\lambda(\cdot))$ be an independent Poisson process with rate $\lambda(\cdot)$. We define the stochastic population model as the sum of Poisson processes in terms of reactions vectors and reaction rates given by
	\bea 
		\bs{X}(t) = \bs{X}(0) + \sum_{i=1}^{n} \bs{R}_i Y_i\left( \int_{0}^{t} \lambda^{i}_{\bstheta}\left( \bs{X}(s) \right) \right). \label{eqn::X PP sum}
	\eea 
	
\begin{figure}[H]
		\centering
		\includegraphics[scale=0.5]{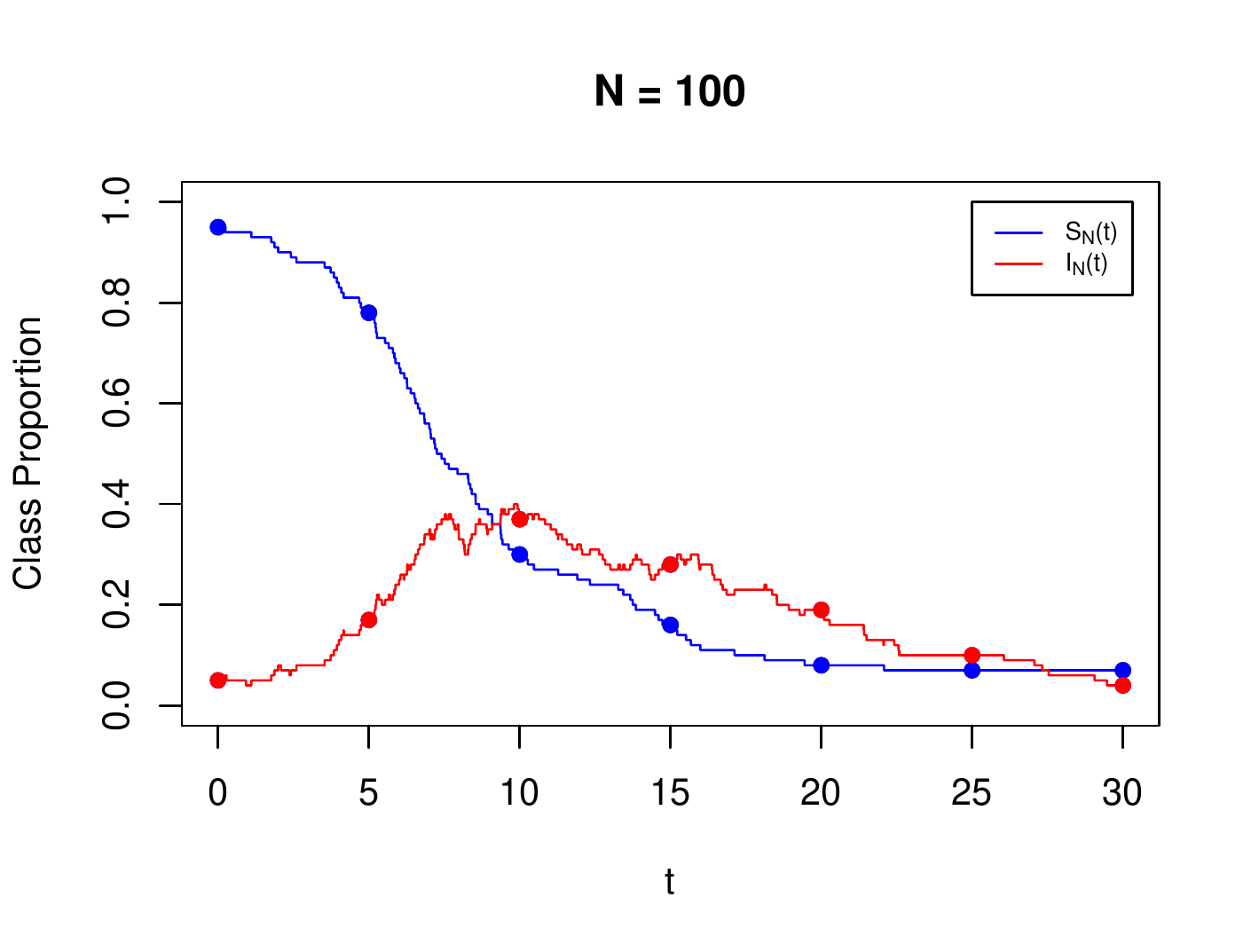}
		\caption{Plot of the infected $I_N(t)$ (red) and susceptible $S_N(t)$ (blue) proportions for a population of size $N = 100$ generated from the stochastic SIR model with $\beta = 0.50$, $\gamma = 0.15$, $I_N(0) = 0.05$, and $S_N(0) = 0.95$.}
		\label{figure::Markov_Pop_SIR}
	\end{figure}

	A classic example of a Markov population model is the stochastic SIR model with a closed population. The SIR model tracks the proportion of susceptible and infected individuals. We let $\bs{X}_N(t) = \left(S_N(t),I_N(t)\right)\p$, denote the scaled proportions of susceptible and infected individuals at time t. There are two possible reactions; a susceptible individual becomes infected $\bs{R}_1 = \left(-1,1\right)\p$, and an infected recovers $\bs{R}_2 = (0,-1)\p$. Susceptible individuals become infected at rate of $\lambda^1_{\bs{\theta}}\left(\bs{X}_N(t)\right) = \beta S_N(t) I_N(t)$, and infected individuals recover at a rate of $\lambda^2_{\bs{\theta}}\left(\bs{X}_N(t)\right) = \gamma I_N(t)$. We note that $\bs{\theta} = (\beta,\gamma)$, where $\beta$ is the contact rate, and $\gamma$ is the recovery rate. For fixed values of $\bs{\theta}$, simulation methods such as the Gillespie algorithm \citep{gillespie1977exact} and tau-leaping \citep{cao2006efficient} can be used to generate stochastic realizations from \eqref{eqn::X PP sum}. The proportions of infected and susceptible individuals from a simulated SIR data set with a population of size $N=100$ are shown in Figure \ref{figure::Markov_Pop_SIR}.  
	
	\section{Diffusion Approximations} \label{section:Diffusion Approximaitons}  

	Markov population model dynamics are controlled by the rate parameters $\bs{\theta}$ which are often unknown and need to be estimated from data. The most common approaches for inference on $\bs{\theta}$ are based on diffusion approximations to \eqref{eqn::X PP sum}. We use the FCLT to construct a system of linear SDEs to help facilitate inference on \eqref{eqn::X PP sum}. Our development follows that of \cite{baxendale2011sustained} and \cite{fricks2018stochastic}.  
	
	Let $\bs{X}_N(t) = \frac{1}{N} \bs{X}(t)$ be the normalized population process. We scale \eqref{eqn::X PP sum} by $N$ to obtain 
	\bea 
		\bs{X}_N(t) = \bs{X}_N(0) + \sum_{i=1}^{n} \bs{R}_i \frac{1}{N} Y_i\left( N \int_{0}^{t} \lambda^{i}_{\bstheta}\left( \bs{X}_N(s) \right) ds \right). \label{eqn::X_N scaled PP} 
	\eea  
	We apply the FCLT for Poisson processes (see Appendix \ref{appendix::FCLT for Poisson Processes}) to each scaled Poisson process in \eqref{eqn::X_N scaled PP} to obtain the Gaussian approximation
	\bea 
		\bs{X}_N(t) \approx \bs{X}_N(t) + \sum_{i=1}^{n} \bs{R}_i \left( \int_{0}^{t} \lambda^{i}_{\bstheta}\left( \bs{X}_N(s) \right) ds + \frac{1}{\sqrt{N}}B_i\left( \int_{0}^{t} \lambda^{i}_{\bstheta}\left( \bs{X}_N(s) \right) ds \right) \right), \label{eqn::X_N approx BM}
	\eea 
	where $B_i(t)$ are independent Brownian motions with variance $t$. We define 
	\bea 
		\mathbb{E}\left[\bs{X}_N(t) | \bs{\theta} \right] &=& \bs{\mu}_{\bs{\theta}}\left(\bs{X}_N(t) | \bstheta \right) = \sum_{i=1}^{n} \bs{R}_i \lambda^{i}_{\bstheta}\left(\bs{X}_N(t)\right), \label{eqn::EM mu} \\
		Cov\left(\bs{X}_N(t) | \bstheta \right) &=& \Sigma_{\bstheta}\left( \bs{X}_N(t) \right) =  \text{G}_{\bstheta}\left(\bs{X}_N(t)\right)\text{G}_{\bstheta}\p\left(\bs{X}_N(t)\right), \label{eqn::EM Cov}
	\eea   
	and differentiate \eqref{eqn::X_N approx BM} to obtain  
	\bea 
		d\bs{X}_N(t) = \bs{\mu}_{\bs{\theta}}\left(\bs{X}_N(t)\right) dt + \text{G}_{\bstheta}\left(\bs{X}_N(t)\right)d\bs{B}(t), \label{eqn::X_N linear time-varying SDE}
	\eea 
	where $d\bs{B}(t) = \left(dB_1(t),dB_2(t),...,dB_d(t)\right)\p$ is a d-dimensional vector of differentiated independent Brownian motions $B_i(t)$. In some cases, the system of SDEs in \eqref{eqn::X_N linear time-varying SDE} yields an analytic solution \citep{oksendal2003stochastic}. However, in most cases, numerical methods are needed to provide approximate solutions. 
	
	The remainder of this work focuses on approximate inference methods for Markov population models that rely on numerical solutions to the system of SDEs in \eqref{eqn::X_N linear time-varying SDE}. In the next section, we highlight the computational bottlenecks of the most commonly used numeric scheme for solving \eqref{eqn::X_N linear time-varying SDE}, Euler-Maruyama. In Section \ref{section::Approximation by Limiting Deterministic Sytstem}, we introduce the JGDLA, which is an approximate joint Gaussian likelihood for \eqref{eqn::X_N linear time-varying SDE} constructed from the solution to a system of ODEs. We then illustrate how to perform inference on \eqref{eqn::X_N linear time-varying SDE} using the JGDLA. 
	
	\section{Euler-Maruyama Approximation} \label{section::Euler-Maruyama Approximation} 
	
	In Section \ref{section:Diffusion Approximaitons} we showed that Markov population models can be approximated by the system of SDEs in \eqref{eqn::X_N linear time-varying SDE}. Performing inference on $\bs{\theta}$ in the absence of an analytic solution consists of two steps; first \eqref{eqn::X_N linear time-varying SDE} is approximated by a numerical method, second an approximate likelihood is constructed from the numerical solution \citep{doucet2009tutorial,kou2012multiresolution}. In this section, we introduce the Euler-Maruyama scheme, which is the most commonly used method for approximating \eqref{eqn::X_N linear time-varying SDE} \citep{sun2015parameter,allen2017primer}. We also highlight the most prominent drawbacks of inference methods that rely on the Euler-Maruyama scheme. 
		
	The Euler-Maruyama scheme approximates the SDE in \eqref{eqn::X_N linear time-varying SDE} with the difference equation
	\bea
		\bs{X}_N(t+\triangle t ) \approx \bs{X}_N(t) + \bs{\mu}_{\bs{\theta}}\left(\bs{X}_N(t) \right) \triangle t +  \sqrt{ \triangle t} \text{G}_{\bstheta}{ \left(\bs{X}_N(t)\right) }  \bs{Z}, \label{eqn::Euler-Maruyama}
	\eea
	where $\bs{Z} = (Z_1,Z_2,...,Z_d)\p \sim N(\bs{0},\mathbb{I}_{d \times d})$ and $\triangle t$ is the time lag between sequential observations of $\bs{X}_N(t)$. From \eqref{eqn::Euler-Maruyama}, we obtain the conditional distributions
	\bea 
		\pi{\left( \bs{X}_N\left(t + \triangle t \right) | \bs{X}_N\left(t\right), \bs{\theta} \right) } \sim N\left(\bs{X}_N(t) + \bs{\mu}_{\bs{\theta}}\left(\bs{X}_N(t) \right) \triangle t, \triangle t \Sigma_{\bstheta}\left( \bs{X}_N(t) \right) \right), \label{eqn::Conditional Euler} 
	\eea 
	which are used to perform inference in a likelihood framework. 
	
	Inference with the Euler-Maruyama scheme generally requires stochastic infill \citep{kou2012multiresolution}. To see this note that the likelihoods in \eqref{eqn::Conditional Euler} require $\bs{X}(t)$ to be observed at all time lags $\triangle t$. Three common reasons Euler-Maruyama requires stochastic infill are; 1) $\bs{X}(t)$ was observed at irregularly spaced time points, 2) the time lag $\triangle t$ is shrunk to improve inference accuracy, and 3) to predict $\bs{X}(t)$ at unobserved time points. We return to the simulated SIR data set from Section \ref{section::Markov Population Models} to illustrate the implications of shrinking $\triangle t$. We assume $\bs{X}(t)$ is observed at time points $t=0,5,10,15,20,25,30$. We could solve the system \eqref{eqn::X_N linear time-varying SDE} with a time step of $\triangle t = 5$, however, we would not be able to predict $\bs{X}(t)$ at any other time points. We consider reducing $\triangle t$ from 5 to 1 to reduce numerical error and predict at the 24 unobserved time points $t=1,2,3,4,6,7,8,9,...,26,27,28,29$. This results in 48 latent states, 24 latent states for each of the two subpopulations tracked by the the SIR model. 
	
	There is no clear way to ignore or integrate over the latent infill states produced by Euler-Maruyama. In the next section, we construct the the JGDLA as an alternative method for approximate inference on Markov population models that does not require stochastic infill for inference or predictions. We show that the JGDLA is a joint Gaussian likelihood for the observed time points of $\bs{X}(t)$ constructed from a deterministic system. The result is a method for approximate inference on Markov population models that relies solely on the solution to a deterministic system.

	\section{JGDLA} \label{section::Approximation by Limiting Deterministic Sytstem} 
	
	In this section we construct the joint Gaussian likelihood of the JGDLA. We follow the results of \cite{kurtz1978strong} to show that the SDE approximation of the Markov population model converges to a system of ODEs. We use the result to construct the JGDLA covariance from the solution to the system of ODEs.
	
	\subsection{Limiting Deterministic Systems} 
	We summarize the results of \cite{kurtz1978strong} which details the sufficient conditions required for the system of SDEs in \eqref{eqn::X_N linear time-varying SDE} to converge to a deterministic system. Assume that $\bs{x}$ is an element in some compact subset $K \subset \mathbb{R}^{d}$ and the following conditions hold:
	\begin{enumerate}
		\item $\sum_{i=1}^{n} |\bs{R}_i| \sup \lambda^{i}_{\bstheta}(\bs{x}) < \infty$,
		\item $\bs{F}(\cdot) = \sum_{i=1}^{n} \bs{R}_i \lambda^{i}_{\bstheta}(\cdot)$ is Lipshitz on $K$,
		\item $\lim\limits_{n\rightarrow \infty} \bs{x}_n = \bs{x}^{\dagger}$.
	\end{enumerate} 
	Theorem 8.1 of \cite{kurtz1981approximation} states that for each $t>0$,
	\bea 
		\lim\limits_{n\rightarrow \infty}\sup_{s\leq t} | \bs{X}_n(t) - \bs{X}^{\dagger}(t)| \rightarrow 0,
	\eea 
	where 
	\bea 
		\bs{X}^{\dagger}(t) = \bs{X}^{\dagger}(0) + \int_{0}^{t} \bs{F}\left(\bs{X}^{\dagger}(s)\right)ds, \label{eqn::X det integral}
	\eea 
	or equivalently 
	\bea 
		d\bs{X}^{\dagger}(t) = \bs{F}(\bs{X}^{\dagger}(t)) =  \sum_{i=1}^{n} \bs{R}_i\lambda^{i}_{\bstheta}\left(\bs{X}^{\dagger}(t)\right)dt, \label{eqn::X det diff}
	\eea
	with initial state $\bs{X}^{\dagger}(0)$. That is, the diffusion approximation of the Markov population process converges to a system of ODEs as the population size $N$ tends towards infinity.
	
	Scientists in fields such as chemistry, ecology, and epidemiology commonly utilize deterministic population models in the form of \eqref{eqn::X det diff} \citep{keeling2011modeling,fricks2018stochastic}. We defined the Markov population model of Section \ref{section::Markov Population Models} in terms of reaction vectors $\bs{R}_i$ and rates $\lambda^i_{\bstheta}(\cdot)$. We note that the reaction vectors and rates of the deterministic system in \eqref{eqn::X det diff} are of the same form as the diffusion approximation in \eqref{eqn::X_N linear time-varying SDE}. This allows stochastic population models to be constructed from their deterministic analogues. In the following sections, we derive the distribution of the JGDLA from the solution of \eqref{eqn::X det diff}.  

	\subsection{Approximate Gaussian Processes} 
	\cite{kurtz1978strong} constructed an approximate distribution for the system of SDEs in \eqref{eqn::X_N linear time-varying SDE} from the limiting deterministic system in \eqref{eqn::X det diff}. We present the results of \cite{kurtz1978strong} required to obtain an approximate distribution for \eqref{eqn::X PP sum}. We begin by considering the difference between $\bs{X}_N(t)$ and its corresponding infinite population limit $\bs{X}^{\dagger}(t)$. Subtracting equations \eqref{eqn::X_N approx BM} and \eqref{eqn::X det integral} gives
	\bea 
		\bs{X}_N(t) - \bs{X}^{\dagger}(t) &\approx& \bs{X}_N(0)-\bs{X}^{\dagger}(0) + \frac{1}{\sqrt{N}} \sum_{i=1}^{n} \bs{R}_i B_i\left(\int_{0}^{t} \lambda^{i}_{\bstheta}{ \left(\bs{X}^{\dagger}(s) \right)} ds \right) \nt \\
		& & + \int_{0}^{t} \bs{F}\left(\bs{X}_N(s)\right) - \bs{F}\left(\bs{X}^{\dagger}(s)\right) ds. \label{eqn::X_N with F} 
	\eea 
	Using a first-order Taylor expansion of $\bs{F}(\bs{X}_N(s))$ about $\bs{F}\left( \bs{X}^{\dagger}(s) \right)$, \eqref{eqn::X_N with F} becomes 
	\bea 
		\bs{X}_N(t) - \bs{X}^{\dagger}(t) &\approx& \bs{X}_N(0)-\bs{X}^{\dagger}(0) + \frac{1}{\sqrt{N}} \sum_{i=1}^{n} \bs{R}_i B_i\left(\int_{0}^{t} \lambda^{i}_{\bstheta}{ \left(\bs{X}^{\dagger}(s) \right) }ds \right) \nt \\ 
		& &  + \int_{0}^{t} \partial \bs{F}\left(\bs{X}^{\dagger}(s)\right)\left(\bs{X}_N(s)-\bs{X}^{\dagger}(s)\right) ds. \label{eqn::X_n with taylor F}
	\eea 
	The FCLT ensures that as the population size $N \rightarrow \infty$, $\sqrt{N}\left(\bs{X}_N(t) - \bs{X}^{\dagger}(t)\right)$ converges in distribution to some zero-mean Gaussian process  
	\bea 
		\bs{V}(t) &=& \int_{0}^{t} \partial \bs{F}\left(\bs{X}^{\dagger}(s)\right)\bs{V}(s)ds + \sum_{i=1}^{n} \bs{R}_i B_i\left(\int_{0}^{t} \lambda^{i}_{\bstheta}\left(\bs{X}^{\dagger}(s)\right) ds \right). \label{eqn::V integral}  
	\eea 
	We define $\text{Q}\left( \bs{X}^{\dagger}(t) \right)$ to be the $d$ by $n$ matrix with its $i^{th}$ column given by $\bs{q}_i\left( \bs{X}^{\dagger}(t) \right) = \bs{R}_i\sqrt{\lambda^{i}_{\bstheta}\left(\bs{X}^{\dagger}(t) \right)}$. We differentiate \eqref{eqn::V integral} to obtain the system of linear SDEs with time-varying coefficients
	\bea 
		d\bs{V}(t) = \partial \bs{F}\left(\bs{X}^{\dagger}(t)\right)\bs{V}(t)dt + \text{Q}\left( \bs{X}^{\dagger}(t) \right) d\bs{B}(t), \label{eqn::dV Linear Sigma} 
	\eea 
	where $d\bs{B}(t) = \left(dB_1(t), dB_2(t),...,dB_d(t) \right)\p$. We then approximate the finite sample process by 
	\bea 
		\bs{X}_N(t) \approx \bs{X}^{\dagger}(t) + \frac{1}{\sqrt{N}} \bs{V}(t). \label{eqn::Approx X by V} 
	\eea  
	
	We note that the approximate distribution in \eqref{eqn::Approx X by V}, which was first constructed by \cite{kurtz1978strong}, is centered about the solution to the system of ODEs $\bs{X}^{\dagger}(t)$ in \eqref{eqn::X det diff} (i.e. $\mathbb{E}\left(\bs{X}_N(t)\right)=\bs{X}^{\dagger}(t)$). From \eqref{eqn::dV Linear Sigma}, we observe that $\bs{V}(t)$ is a zero-mean Gaussian process, whose covariance depends only on the deterministic solution $\bs{X}^{\dagger}(t)$. All past work on this approximation has focused on simulation rather than inference \citep{kurtz1981approximation,baxendale2011sustained,fricks2018stochastic}. In this work we develop methods for statistical inference on $\bs{\theta}$ using the approximation \eqref{eqn::Approx X by V}. To this end, we construct the covariance for the JGDLA as a function of the solution to the deterministic system in \eqref{eqn::X det diff}. To our knowledge, we are the first to construct this covariance and use it for statistical inference. 
	
	\subsection{Deterministic Covariance Structures} \label{section::Cov}
	Here we define the joint Gaussian likelihood of the JGDLA. To do so, we first solve for the covariance of $\bs{V}(t)$ by solving \eqref{eqn::dV Linear Sigma}. While this covariance cannot, in general, be obtained analytically, we develop a novel form for the covariance which can easily be approximated numerically. To do so, we propose a separable solution of the form 
	 \bea 
	 	\bs{V}(t) = U(t)\bs{Y}(t), \bs{V}(0) = \bs{0}, \label{eqn::V seperable}
	 \eea 
	 where $U(t)$ is a $d$ by $d$ matrix and $\bs{Y}(t)$ is a $d$-dimensional Gaussian process. The solution to \eqref{eqn::V seperable} (see Appendix \ref{appendix::Solving for V(t)} for details) requires 
	 \bea 
	 	dU(t) &=& \partial \bs{F}\left(\bs{X}^{\dagger}(t)\right)U(t), U(0) = \mathbb{I}_{d \times d}, \label{eqn::dU} \\
	 	\bs{Y}(t) &=& \int_{0}^{t} U^{-1}(s)\text{Q}\left( \bs{X}^{\dagger}(s) \right) d\bs{B}(s), \bs{Y}(0) = \bs{0}, \label{eqn::Y}
	 \eea 
	 where $\partial \bs{F}\left(\bs{X}^{\dagger}(t)\right)$ and $Q\left(\bs{X}^{\dagger}(t)\right)$ are defined in \eqref{eqn::dV Linear Sigma}, and are respectively a $d$ by $d$ matrix of partial derivatives of $\bs{F}\left(\bs{X}^{\dagger}(t)\right)$ and a $d$ by $n$ matrix with columns $\bs{q}_i\left(\bs{X}^{\dagger}(t)\right) = \bs{R}_i \sqrt{\lambda_{\bs{\theta}}^i\left(\bs{X}^{\dagger}(t)\right)}$. An exact solution to \eqref{eqn::dU} does not exist unless $\partial \bs{F}\left(\bs{X}^{\dagger}(t)\right)$ is constant or commutes. For cases in which $\partial \bs{F}\left(\bs{X}^{\dagger}(t)\right)$ is not constant and does not commute, \eqref{eqn::dU} is a system of time inhomogeneous linear differential equations that must be solved numerically. We obtain an approximate solution to \eqref{eqn::dU} by first solving the deterministic system in \eqref{eqn::X det diff} for $\bs{X}^{\dagger}(t)$, then numerically solving for $U(t)$ in \eqref{eqn::dU}.
	 
	 Since $\bs{V}(t)$ is a zero-mean Gaussian process, the covariance of $\bs{V}(t)$ is a linear transformation of the covariance of $\bs{Y}(t)$. We define the covariance of the zero-mean process $\bs{Y}(t)$ and obtain the covariance of $\bs{V}(t)$ by matrix multiplication. For each $i = 1,2,...,n$, we define the $d$-dimensional vectors $\bs{a}_i(t) = \left( a_{i1}(t),a_{i2}(t),...,a_{id}(t) \right)\p = U^{-1}(t)\bs{R}_i$. Then from \eqref{eqn::Y}, each component of $\bs{Y}(t)$ is 
	 \bea 
	 	Y_j(t) = \sum_{i=1}^{n} \int_{0}^{t} a_{ij}(s) \sqrt{\lambda^{i}_{\bstheta}\left( \bs{X}^{\dagger}(s) \right) } dB_i(s). \label{eqn::Y_j}
	 \eea 
	 It follows from \eqref{eqn::Y_j} that 
	 \begin{equation}
		 Cov\left( Y_j(t),Y_k(t) \right) = \begin{cases}
		 \sum_{i=1}^{n} \int_{0}^{t} a_{ij}^2(s) \lambda^{i}_{\bstheta}\left( \bs{X}^{\dagger}(s) \right)ds, & j = k \\
		 \sum_{i=1}^{n} \int_{0}^{t} a_{ij}(s)a_{ik}(s) \lambda^{i}_{\bstheta}\left( \bs{X}^{\dagger}(s) \right)ds, & j \neq k 
		 \end{cases}. \label{eqn::Cov Y} 
	 \end{equation}
	 The covariance of $\bs{V}(t)$ is then given by
	 \bea 
	 	Cov(\bs{V}(t)) = Cov(U(t)\bs{Y}(t)) = U(t) Cov(\bs{Y}(t)) U\p(t). \label{eqn::Cov V} 
	 \eea 
	 It follows from \eqref{eqn::Cov V} that for $s < t$, 
	 \bea 
	 	Cov\left( \bs{X}_N(s), \bs{X}_N(t) \right)	= \frac{1}{N} U(s) Cov(\bs{Y}(s)) U\p(t). \label{eqn::Cov X_N}
	 \eea 
	 We use the approximation in \eqref{eqn::Approx X by V} and \eqref{eqn::Cov X_N} to construct a joint Gaussian likelihood for observed time points $t_1,t_2,t_3,..,t_T$. Define observations of the Markov population model $\bs{X}_N = \left(\bs{X}_N(t_1),\bs{X}_N(t_2),...,\bs{X}_N(t_T)\right)\p$ and the ODE solution for a given $\bs{\theta}$ as $\bs{X}^{\dagger} = \left(\bs{X}^{\dagger}(t_1),\bs{X}^{\dagger}(t_2),...,\bs{X}^{\dagger}(t_T)\right)\p$. Let 
	 \begin{equation}
	 	\Sigma_Y = \begin{bmatrix}
	 		Cov(\bs{Y}(t_1)) & Cov(\bs{Y}(t_2)) & ... & Cov(\bs{Y}(t_T)) \\
	 		Cov(\bs{Y}(t_1)) & Cov(\bs{Y}(t_2)) & ... & Cov(\bs{Y}(t_T)) \\ 
	 		\vdots & \vdots & ... & \vdots \\
	 		Cov(\bs{Y}(t_1)) & Cov(\bs{Y}(t_2)) & ... & Cov(\bs{Y}(t_T))
	 	\end{bmatrix}, \nt 
	 \end{equation}
	 and 
	 \begin{equation}
	 	U = \begin{bmatrix}
	 		U(t_1) & U(t_1) & ... & U(t_1) \\
	 		U(t_2) & U(t_2) & ... & U(t_2) \\ 
	 		\vdots & \vdots & ... & \vdots \\ 
	 		U(t_T) & U(t_T) & ... & U(t_T)  
	 	\end{bmatrix}. \nt 
	 \end{equation} 
	 The resulting JGDLA covariance is
	 \begin{equation}
	 	\Sigma^{\dagger}_{\bs{\theta}} = \frac{1}{N} U * \Sigma_Y * U\p, \label{eqn::Sigma Dagger}
	 \end{equation}
	 where $``*"$ denotes element-wise multiplication. Our novel JGDLA likelihood is then given by  
	 \begin{equation} 
	 	\pi\left(\bs{X}_N|\bs{\theta},\bs{X}^{\dagger} \right) \sim N\left(\bs{X}^{\dagger}, \Sigma^{\dagger}_{\bs{\theta}} \right). \label{eqn::JGDLA likelihood}
	\end{equation} 
	From \eqref{eqn::JGDLA likelihood}, we see that JGDLA likelihood evaluations rely solely on solutions to a deterministic system. In summary, the JGDLA is obtained by the following procedure. For a set of parameters $\bs{\theta}$, 
	 \begin{enumerate}
	 	\item Solve the system of ODEs in \eqref{eqn::X det diff} for $\bs{X}^{\dagger}(t)$.
	 	\item Solve equation \eqref{eqn::dU} for $U(t)$.
	 	\item Solve the integrals in \eqref{eqn::Cov Y} at all observed time points to construct $\Sigma_Y$. 
	 	\item Construct $\Sigma^{\dagger}_{\bs{\theta}}$ in \eqref{eqn::Sigma Dagger}.
	 	\item Evaluate joint likelihood $\pi\left(\bs{X}_N|\bs{\theta},\bs{X}^{\dagger} \right)$ given in \eqref{eqn::JGDLA likelihood}.
	 \end{enumerate}
 
	We note that the integrals involved in the algorithm described above are all deterministic. The integrals are solved numerically, but do not require stochastic infill. Further, predictions at unobserved time points can be obtained via conditional predictive normal distributions. Thus, the JGDLA provides a framework for inference based on the approximate joint likelihood \eqref{eqn::JGDLA likelihood} of all observed data which relies solely on the solution to a deterministic system. 
	  
	\section{Simulation Study} \label{section::Simulation Study}
	
	In this section we formulate the stochastic SIR model directly from its popular deterministic analogue. We show that the JGDLA offerers improved predictive power in comparison to the Euler-Maruyama scheme of Section \ref{section::Euler-Maruyama Approximation}.

	\subsection{The Stochastic SIR Model}
	
	Deterministic SIR population models are widely used in epidemiology to model disease outbreaks \citep{keeling2011modeling}. The SIR model tracks the deterministic proportions of susceptible $S^{\dagger}(t)$, infected $I^{\dagger}(t)$, and recovered $R^{\dagger}(t)$ individuals in a closed population of size $N$. In large populations, systems of ODEs are commonly used to model disease dynamics \citep{keeling2011modeling}. The differential equations governing the deterministic system are given by 
	\bea 
		\frac{d}{dt} S^{\dagger}(t) &=& -\beta S^{\dagger}(t)I^{\dagger}(t), \label{eqn::S deterministic} \\
		\frac{d}{dt} I^{\dagger}(t) &=& \beta  S^{\dagger}(t)I^{\dagger}(t) - \gamma I^{\dagger}(t), \label{eqn::I deterministic} \\
		\frac{d}{dt} R^{\dagger}(t) &=& \gamma I^{\dagger}(t), \label{eqn::R deterministic}
	\eea 
	where $1 = I^{\dagger}(t) + S^{\dagger}(t) + R^{\dagger}(t)$ due to the assumption of a constant population size. Note that since $R^{\dagger}(t) = 1 - I^{\dagger}(t) - S^{\dagger}(t)$, the system can be reduced to equations \eqref{eqn::S deterministic} and \eqref{eqn::I deterministic}. The two unknown parameters $\bs{\theta} = (\beta,\gamma)$, are the direct transmission rate $\beta \in (0,\infty)$ and the recovery rate $\gamma \in (0,\infty)$. 
	
	We construct the stochastic SIR model from the system of ODEs by defining the reaction vectors and their corresponding deterministic rates. Recall the two reactions that may occur previously defined in Section \ref{section::Markov Population Models}, a susceptible individual becomes infected $\bs{R}_1 = (-1, 1)\p$, or an infected recovers $\bs{R}_2 = (0, -1)\p$. We denote the deterministic class proportions $\bs{X}^{\dagger}(t) = \left(S^{\dagger}(t), I^{\dagger}(t)\right)\p$, and define the reaction rates 
	\bea 
		\lambda^{1}_{\bstheta}\left(\bs{X}^{\dagger}(t)\right) &=& \beta S^{\dagger}(t) I^{\dagger}(t), \label{eqn::Lambda_1 SIR} \\ 
		\lambda^{2}_{\bstheta}\left(\bs{X}^{\dagger}(t)\right) &=& \gamma I^{\dagger}(t). \label{eqn::Lambda_2 SIR} 
	\eea 
	We let $\bs{X}_N(t) = \left(S_N(t),I_N(t)\right)\p$ denote the random proportions of susceptible and infected individuals at time $t$. We obtain the stochastic SIR model by swapping $\bs{X}_N(t)$ and $\bs{X}^{\dagger}(t)$ in (\ref{eqn::Lambda_1 SIR}--\ref{eqn::Lambda_2 SIR}). 
	
\begin{figure}[H]
		\centering
		\includegraphics{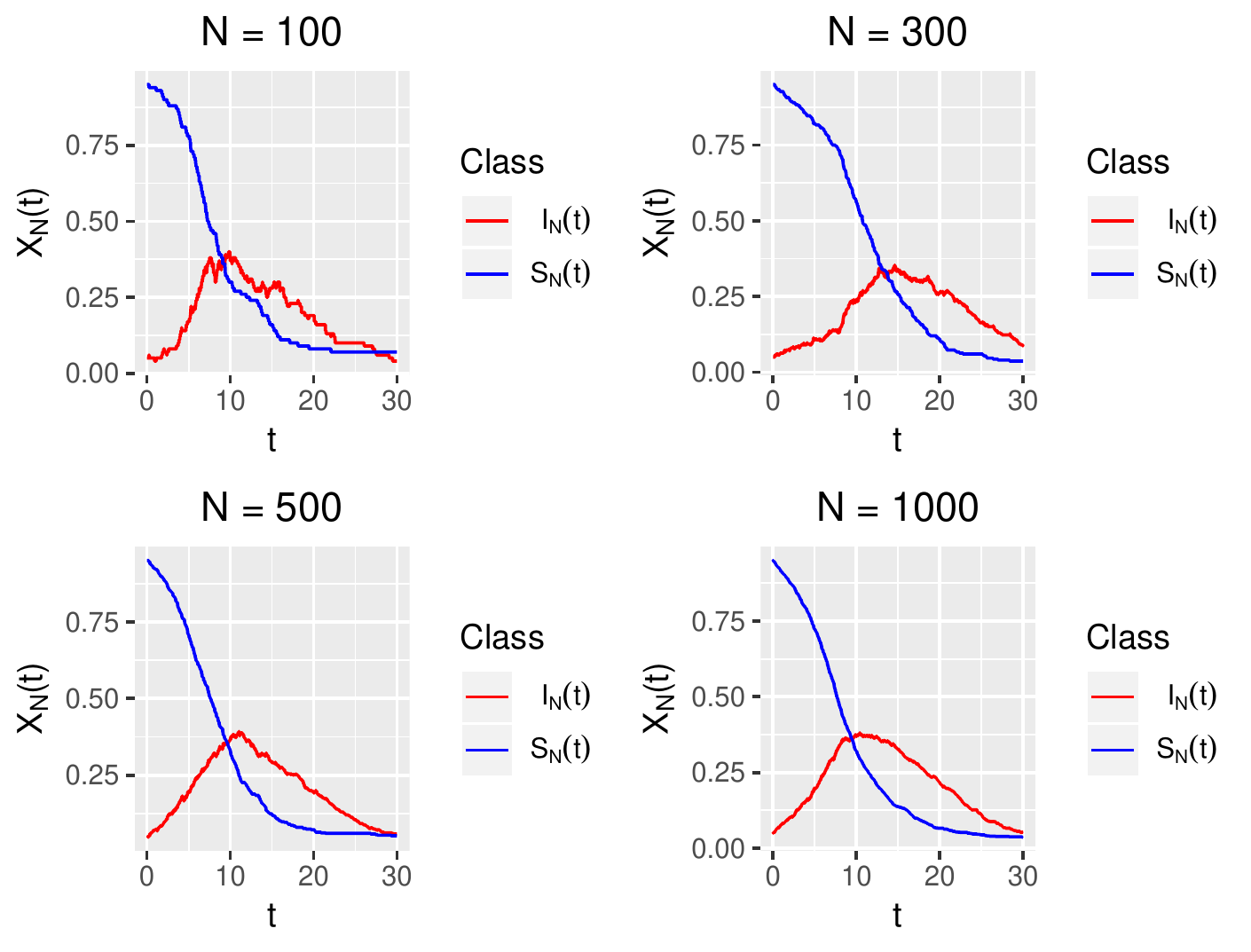}
		\caption{Plot of the infected $I_N(t)$ (red) and susceptible $S_N(t)$ (blue) proportions for $N = 100, 300, 500$, and $1000$ generated from a stochastic SIR model with $\beta = 0.5, \gamma = 0.15, S_N(0) = 0.95$, and $I_N(0)=0.05$.}
		\label{figure::SIR_Sim_Data}
	\end{figure}

	We use the Gillespie algorithm \citep{gillespie1977exact} to generate four datasets with $\beta = 0.50, \gamma = 0.15,$ and initial conditions $S_N(0) = 0.95$, and $I_N(0) = 0.05$ for population sizes $N = 100, 300, 500, 1000$ over the time interval $(0,30]$ (see Figure \ref{figure::SIR_Sim_Data}). We denote the observed times as $T^{obs} = \{5,10,15,20,25,30\}$. We will compare the models in terms on mean absolute prediction error (MAPE) on the proportions of infected for each approximation method at the infill time points $T^{pred} = \{1,2,3,4,6,7,8,9,...,26,27,28,29\}$. Note that our partition is denoted $t_0 = 0, t_1 = 1, t_2 = 2,...,t_{30} = 30$.
	
	\subsection{Models and Model Fitting} 
	
	We first fit the simulated data to the fully deterministic system as a point of comparison for the stochastic models. The fully deterministic model is fit with likelihood given by 
	\bea 
		\pi{ \left(\bs{X}^{obs}_N(t) | \bs{X}^{\dagger}(t),\sigma \right)} \sim N\left( \bs{X}^{\dagger}(t), \sigma^2 \mathbb{I}_{2 \times 2} \right). \label{eqn::Sim Study Det Model} 
	\eea 
    Maximum likelihood estimates from the deterministic SIR model likelihood in \eqref{eqn::Sim Study Det Model}, with mean $\bs{X}^{\dagger}(t)$ being the numerical optimization solution of \eqref{eqn::X det diff} and a function of $\bs{\theta} = (\beta,\gamma)$, which are estimated. We use \textit{ode} from the \textit{deSolve} package in \textit{R} with a time step of 0.1 to solve for $\bs{X}^{\dagger}(t)$ and obtain maximum likelihood estimates for $(\sigma,\beta,\gamma)$ using the built-in optimizer \textit{optim} in $R$.    
	
	We follow the summary procedure in Section \ref{section::Cov} to obtain the JGDLA likelihood for $\pi{\left(\bs{X}^{obs}_N|\bs{X}^{\dagger},\bs{\theta}\right)}$, where $\bs{X}^{obs}_N = \left(\bs{X}^{obs}_N(5),\bs{X}^{obs}_N(10),...,\bs{X}^{obs}_N(30)\right)\p$ and $\bs{X}^{\dagger} = \left(\bs{X}^{\dagger}(5),\bs{X}^{\dagger}(10),...,\bs{X}^{\dagger}(30)\right)\p$. We use \textit{ode} from the \textit{deSolve} package in \textit{R} with a time step of 0.1 to solve all the integrals required to form the JGDLA likelihood. We again obtain maximum likelihood estimates for the JGDLA model \textit{optim} in $R$. Details on the construction of the JGDLA are contained in Appendix \ref{appendix::JGDLA for SIR}. 
	
 	For comparison, we also make inference based on an Euler-Maruyama approximation. We implement the Euler-Maruyama scheme (EM) for the stochastic SIR model with a time-lag of $\triangle t = 1$ by defining
	\[
		\text{G}_{\bs{\theta}}\left( \bs{X}_N(t) \right) = \begin{bmatrix}
		\sqrt{ \frac{\lambda^{1}_{\bstheta}\left(\bs{X}_N(t)\right)}{N} } & 0 \\
		-\sqrt{ \frac{\lambda^{1}_{\bstheta}\left(\bs{X}_N(t)\right)}{N} } & \sqrt{ \frac{\lambda^{1}_{\bstheta}\left(\bs{X}_N(t)\right)}{N}+ \frac{\lambda^2_{\bstheta}\left(\bs{X}_N(t)\right)}{N} }
		\end{bmatrix}, 
	\]
	and $\bs{\mu}_{\bs{\theta}}\left( \bs{X}_N(t) \right) = \left(-\lambda^{1}_{\bstheta}\left( \bs{X}_N(t) \right), \lambda^{1}_{\bstheta}\left( \bs{X}_N(t) \right) - \lambda^2_{\bstheta}\left(\bs{X}_N(t)\right) \right)'$ in \eqref{eqn::Conditional Euler} of Section \ref{section::Euler-Maruyama Approximation}. We also consider the independent Euler-Maruyama (EM Ind) model, which is obtained by setting the off-diagonal elements in $\text{G}_{\bs{\theta}}\left( \bs{X}_N(t) \right) $ of \eqref{eqn::Conditional Euler} to be zero. Taking $\text{G}_{\bs{\theta}}\left( \bs{X}_N(t) \right)$ to be diagonal allows for fast inversions and likelihood evaluations, and is common in some settings \citep{wikle2010general}.
 	
	Both Euler-Maruyama schemes require 48 latent states to be estimated at $T^{pred}$. We fit both Euler-Maruyama schemes via MCMC. We specify $iid$ half normal scale 1 priors for $\pi{\left( \beta\right)}$ and $\pi{\left( \gamma \right) }$, giving $\pi{\left(\bs{\theta}\right)} = \pi{\left( \beta\right)} \pi{\left( \gamma \right) }$. Let $\bs{X}^{pred}$ denote the collection of $\bs{X}_N(t)$ at infill time points $T^{pred}$. We estimate the latent infill states by drawing all-at-once Metropolis-Hasting samples from the conditional distribution 
	\bea 
		 \pi{\left( \bs{X}^{pred},\bs{\theta}|\bs{X},\bs{X}_N(0)\right)} \propto \prod_{i=1}^{30} \pi{\left( \bs{X}_N\left( t_{i}\right) | \bs{X}_N\left( t_{i-1}\right), \bs{\theta}  \right)} \pi{\left(\bs{\theta}\right)}, \label{eqn::conditional posterior Euler}
	\eea
	where $\pi{\left( \bs{X}_N\left( t_{i}\right) | \bs{X}_N\left( t_{i-1}\right), \bs{\theta}  \right)}$ is the conditional normal likelihood obtained from the Euler-Maruyama scheme in \eqref{eqn::Conditional Euler}. We draw 300,000 burn-in states from \eqref{eqn::conditional posterior Euler} to allow for the Markov chain to reach a stationary distribution and store an additional 100,000 post burn-in states to perform the analysis.

	\subsection{Simulation Study Discussion} 
	
	These three approaches to inference are all compared in terms of mean absolute prediction error (MAPE) on the predicted infected. The MAPE at the 24 infill time points for the infected is computed as $MAPE = \sum_{t \in T^{pred}} | I_N(t) - I^{truth}_N(t)| / 24 $. To compute the MAPE for the JGDLA, we take the expectation of the conditional predictive multivariate normal distribution for $\bs{X}^{pred}$ conditioned on $\bs{X}^{obs}_N$ and the maximum likelihood estimates for $\bstheta$. For the Euler-Maruyama schemes, we compute the MAPE as $\sum_{t \in T^{pred}} \left(\sum_{l=1}^{L}   |I_N^{(l)}(t) - I_N^{true}(t)| / L\right)/24$, where $\bs{X}^{(l)}(t) = \left( S_N^{(l)}(t), I_N^{(l)}(t) \right)\p$ represents a post burn-in Metropolis hasting sample from the posterior infill distribution. The MAPE for the ODE fit takes the ODE solution $\bs{X}^{\dagger}(t)$ for the maximum likelihood estimates of $\bs{\theta}$ as the predicted value at the infill locations. 
	
	We summarize the MAPE for each approximation method in Table \ref{table::Simulation Study MAPE} and note that all 95\% credible/confidence intervals contained the true value of $\bs{\theta}$. We see that the JGDLA offers improved predictive power in comparison to both Euler-Maruyama approximations and the deterministic fits. We note that the Euler-Maruyama scheme can be improved by refining the time discretization. However, lattice refinement requires all latent states at new infill locations to be inferred. For example, the Euler-Maruyama scheme required 48 latent states to perform inference with $\triangle t = 1$ in the SIR example. No stochastic infill was required to obtain the JGDLA model fit. The results of this study suggest that the JGDLA offers an improved model fit in comparison to the Euler-Maruyama scheme considered, and does not require latent states to be stochastically estimated. Inference under the JGDLA is straightforward, as it allows for direct evaluation of the joint likelihood of the data. 
		
	\begin{table}[H]
		\centering 
		\begin{tabular}{lllll}
			N    & EM    &  EM Ind  &  JGDLA & ODE \\
			\hline
			$100$ & 0.02389 & 0.02498 & \textbf{0.01464} & 0.01967\\ 
			$300$ & 0.01484 & 0.01627 & \textbf{0.01252} & 0.02327 \\ 
			$500$ & 0.00863 & 0.00968 & \textbf{0.00612} & 0.00678 \\ 
			$1,000$ & 0.00679 & 0.00758 & \textbf{0.00456} & 0.00868               
		\end{tabular}
		\caption{MAPE for the predicted infected proportions for the four models; Euler-Maruyama (EM), Euler-Maruyama with off-diagonal covariance terms set to 0 (EM Ind), JGDLA, and the ODE system fitted to the four simulated data sets with $N=100,300,500,1000$.}
		\label{table::Simulation Study MAPE} 
	\end{table}
	
	\section{Data Analysis: COVID Cruise Ship} \label{section::Covid Data Example}
	
	On 5 February 2020 a cruise ship hosting 3711 people docked for a 2-week quarantine in Yokohama, Japan after a passenger tested positive for the coronavirus disease (COVID-19) \citep{mizumoto2020estimating}. Random testing of passengers and crew members began on February $5^{th}$ and continued through February $20^{th}$. We note that on February $11^{th}$ and $14^{th}$ no testing occurred and denote the fourteen observed times $T^{obs} = \{1,2,3,..,6,8,9,11,12,...,16\}$. The number of tests administered per day $n_t$, positive tests per day $y_t$, and total number of individuals remaining on the ship are shown in Table \ref{table::Covid Data}. Once exposed to COVID, susceptible individuals experience an incubation prior to becoming infectious \citep{chen2020mathematical}. To account for the incubation period, we fit a stochastic susceptible-exposed-infected-removed (SEIR) model to the Princess Diamond cruise ship COVID-19 outbreak data set. We use the JGDLA to form a joint distribution for the latent states of the SEIR model. 
	
	We specify the system of ODEs governing the SEIR model 
	\bea 
		\frac{d}{dt}S^{\dagger}(t) &=& -\beta I^{\dagger}(t)S^{\dagger}(t)  - \mu_S(t)S^{\dagger}(t), \nt \\
		\frac{d}{dt}E^{\dagger}(t) &=& \beta I^{\dagger}(t)S^{\dagger}(t) - \alpha E^{\dagger}(t), \nt \\ 
		\frac{d}{dt}I^{\dagger}(t) &=& \alpha E^{\dagger}(t) - \gamma I^{\dagger}(t), \nt \\
		\frac{d}{dt}R^{\dagger}(t) &=& \gamma I^{\dagger}(t)  + \mu_S(t) S^{\dagger}(t), \nt 
	\eea 
	where $1=I^{\dagger}(t)+S^{\dagger}(t)+E^{\dagger}(t)+R^{\dagger}(t)$. We note that the system contains three compartments $\bs{X}^{\dagger} = \left(S^{\dagger}(t), E^{\dagger}(t), I^{\dagger}(t)\right)\p$, since $R^{\dagger}(t) = 1 - S^{\dagger}(t) - E^{\dagger}(t) - I^{\dagger}(t)$. The contact rate $\beta$, incubation rate $\alpha$, and the recovery rate $\gamma$ all have support on the positive reals. Let $\bs{X}^{\dagger} = \left(\bs{X}^{\dagger}(1),...,\bs{X}^{\dagger}(6),\bs{X}^{\dagger}(8),\bs{X}^{\dagger}(9),\bs{X}^{\dagger}(11),...\bs{X}^{\dagger}(16)\right)\p$ denote the ODE solution at times $T^{obs}$. We build a joint distribution for the latent proportions $\bs{X}_N(t) = \left( S_N(t), E_N(t), I_N(t) \right)\p$ using JGDLA at the fourteen observed time points $\bs{X}_N = \left(\bs{X}_N(1),...,\bs{X}_N(6),\bs{X}_N(8),\bs{X}_N(9),\bs{X}_N(11),...\bs{X}_N(16)\right)\p$ centered at $\bs{X}^{\dagger}$.

	We estimate $\bs{\theta} = (\beta,\alpha,\gamma)$ and the initial conditions $\bs{X}_N(0) = \left( S_N(0), E_N(0), I_N(0) \right)\p$, where $t=0$ denotes February $4^{th}$. We account for the disembarkment of susceptible passengers with the inclusion of a fixed time-varying rate $\alpha_S(t)$ estimated from passenger records. We model the number of observed seropositive individuals on day \textit{t} as 
	\bea 
		y_t \sim Binom\left(n_t, P(t) = \frac{I_N(t)}{I_N(t)+S_N(t)+E_N(t)} \right). \label{eqn::Covid Likelihood}
	\eea 
	We obtain the stochastic model from the deterministic SEIR by defining the four reaction vectors; a susceptible becomes exposed $\bs{R}_1 = \left(-1, 1, 0 \right)\p$, a susceptible disembarks from the ship $\bs{R}_2 = (-1,0,0)\p$, an exposed individual becomes infected $\bs{R}_3 = (0,-1,1)\p$, or an infected individual recovers $\bs{R}_4 = (0,0,-1)\p$. We define the corresponding reaction rates $\lambda^1_{\bstheta}\left(\bs{X}^{\dagger}(t)\right) = \beta I^{\dagger}(t)S^{\dagger}(t)$, $\lambda^2_{\bstheta}\left( \bs{X}^{\dagger}(t) \right) = \mu_S(t) S^{\dagger}(t)$, $\lambda^3_{\bstheta}\left( \bs{X}^{\dagger}(t) \right) = \alpha E^{\dagger}(t)$, and $\lambda^4_{\bstheta}\left( \bs{X}^{\dagger}(t) \right) = \gamma I^{\dagger}(t)$.  
	
	We elect to take a Bayesian approach to inference and fit the model via MCMC. We place priors of $\beta,\alpha \stackrel{iid}{\sim} TN_{(0,\infty)}\left(0,15^2\right)$, $\gamma \sim TN_{(0,\infty)}\left(0,0.3^2\right)$, $S_N(0) \sim TN_{(0,1)}\left(0,0.3^2\right)$, and $I_N(0) \sim TN_{(0,1)}\left(0,0.1^2\right)$, where $TN_{(a,b)}(c,d)$ denotes a truncated normal distribution with support $(a,b)$, center $c$, and scale parameter $d$. We draw 100,000 all-at-once Metropolis Hastings samples of $\bs{\theta}$ and $\bs{X}_N(0)$. We discard 10,000 samples as burn-in states and use the remaining 90,000 samples to perform the analysis.   
	
	The likelihood in \eqref{eqn::Covid Likelihood} must be approximated due to its dependence on the latent states $\bs{X}_N$.  We perform a Monte Carlo approximation of \eqref{eqn::Covid Likelihood} by drawing 1,000 samples from the JGDLA joint distribution $\bs{X}_N^{(l)} \sim \pi{\left( \bs{X}_N | \bs{\theta},\bs{X}^{\dagger} \right)}$, assigning $P^{(l)}(t) = I^{(l)}_N(t) / \left( I^{(l)}_N(t)+S^{(l)}_N(t)+E^{(l)}_N(t) \right)$ from the JGDLA samples, then approximating the likelihood 
	\bea 
		\hat{\pi}{\left(\bs{y} | \bs{\theta} \right)} = \frac{1}{1000}\sum_{l=1}^{1000} \prod_{t \in T^{obs}}\pi{\left( y_t | P^{(l)}(t) \right)}.   \label{eqn::MC Likelihood}
	\eea 
	We use the approximate likelihood in \eqref{eqn::MC Likelihood} to draw all-at-once Metropolis-Hasting samples from the posterior conditional distribution $\pi{\left( \bs{\theta},\bs{X}_N(0) | \bs{y} \right)} \propto \hat{\pi}{\left(\bs{y} | \bs{\theta} \right)}\pi{\left( \bs{X}_N(0) \right)}\pi{\left(\bs{\theta}\right)}$, where $\bs{y}$ is the vector of $y_t$ at $T^{obs}$. Further details on model fitting are included in Appendix \ref{appendix::JGDLA for SEIR}. 
	
	From Figure \ref{figure::Covid_Data_Fit}, we see that the posterior mean estimate for the probability of being infected captures the mean of the observed proportions well. All parameter estimates are summarized in Table \ref{table::Covid Param Ests}. We note that $\alpha$ is the rate at which individuals move from exposed to infected. Studies have found that symptoms take roughly five to six days to appear \citep{chen2020mathematical}. Our estimate is most likely lower due to the fact that several tests were administered to a small population (i.e. many were tested positive before symptoms began). We also note that the recovery/removal rate is roughly 1.14 days. This estimate is lower than the known time to recover from SARS-COV2, as individuals on the ship were not tested once found positive, and likely confined to their rooms (i.e. quarantined), and are thus functionally removed from the population, even though they are still infectious. Our stochastic SEIR model, assisted by the use of the JGDLA, allows us to understand the mean behavior of the small population of the Princess Diamond cruise ship.

	\begin{table}[H]
		\centering 
		
		\begin{tabular}{lllll}
			t  & Date (2020) & Number of Tests ($n_t$) & Positive Tests $(y_t)$ & On Ship \\
			\hline
			1  & 5 Feb       & 31                      & 10                   & 3711    \\
			2  & 6 Feb       & 71                      & 10                   & 3711    \\
			3  & 7 Feb       & 171                     & 41                   & 3711    \\
			4  & 8 Feb       & 6                       & 3                    & 3711    \\
			5  & 9 Feb       & 57                      & 6                    & 3711    \\
			6  & 10 Feb      & 103                     & 65                   & 3711    \\
			7  & 11 Feb      & \textit{NA}             & \textit{NA}           & 3711    \\
			8  & 12 Feb      & 53                      & 39                   & 3711    \\
			9  & 13 Feb      & 221                     & 44                   & 3711    \\
			10 & 14 Feb      &\textit{NA}              & \textit{NA}           & 3451    \\
			11 & 15 Feb      & 217                     & 67                   & 3451    \\
			12 & 16 Feb      & 289                     & 70                   & 3451    \\
			13 & 17 Feb      & 504                     & 99                   & 3183    \\
			14 & 18 Feb      & 681                     & 88                   & 3183    \\
			15 & 19 Feb      & 607                     & 79                   & 3183    \\
			16 & 20 Feb      & 52                      & 13                   & 2213   
		\end{tabular} 
	\caption{Covid data}
	\label{table::Covid Data}
	\end{table}
	
	\begin{table}[H]
		\centering 
		\begin{tabular}{lll}
			Parameter & Posterior Mean Estimate & 95\% CI \\
			\hline
			$\beta$ & 3.108  & (1.433,5.534) \\
			$\alpha$ & 0.526 & (0.422,0.691) \\
			$\gamma$ & 0.876 & (0.605,1.172) \\
			$S_N(0)$ & 0.545 & (0.265,0.754) \\
			$I_N(0)$ & 0.088 & (0.008,0.193) 
		\end{tabular} 
	\caption{Posterior mean estimates and 95\% credible intervals for estimated SEIR model parameters.}
	\label{table::Covid Param Ests}
	\end{table}

	\begin{figure}[H]
		\centering
		\includegraphics{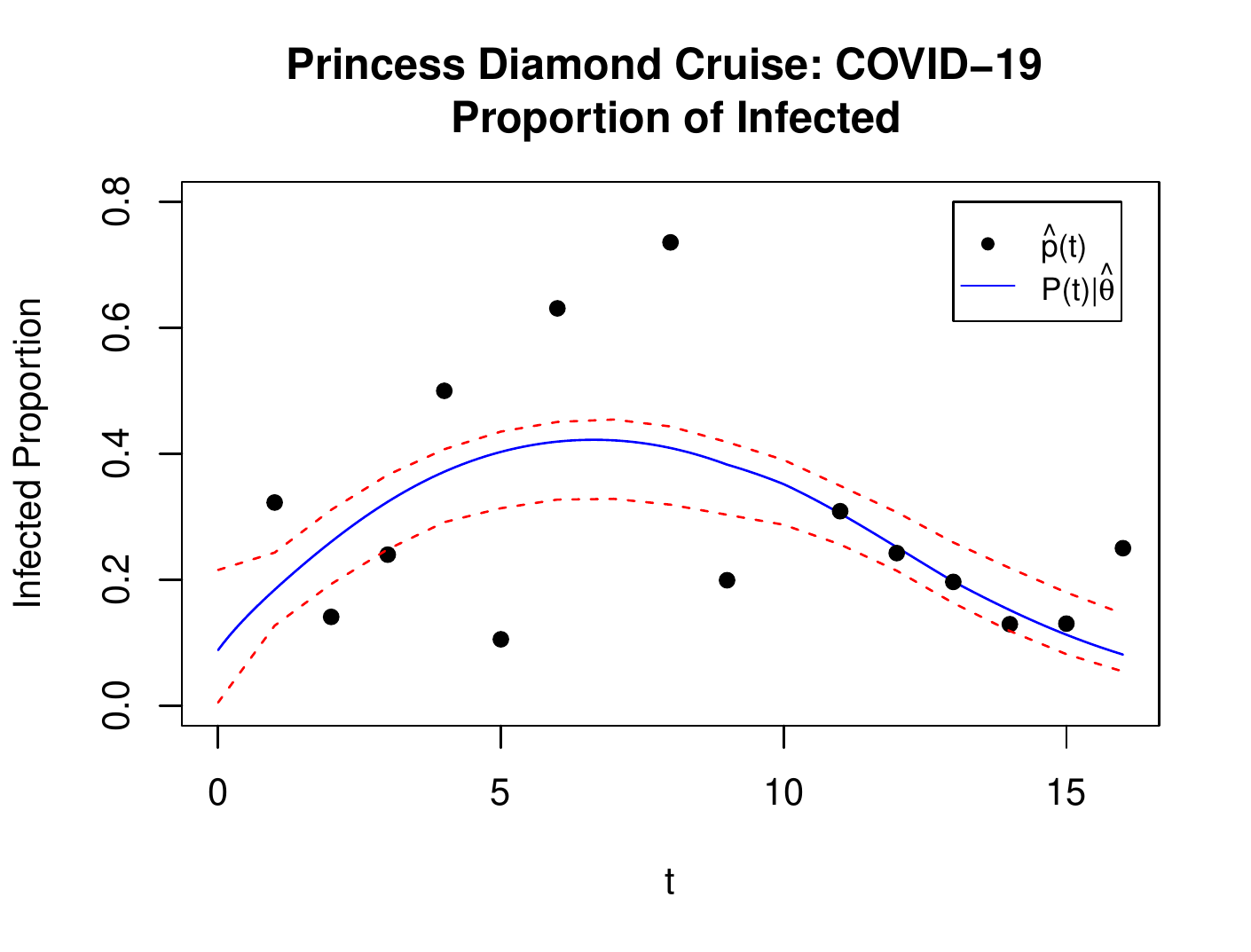}
		\caption{Plot of Princess Diamond Cruise COVID-19 proportions of infected. Sample proportions estimated from Table \ref{table::Covid Data} are plotted in black. The deterministic curve for $P(t)$ fitted from the posterior mean estimate of $\bs{\theta}$ is shown in blue, with 95\% credible intervals in red.}
		\label{figure::Covid_Data_Fit}
	\end{figure}

	\section{Discussion} \label{section::Discussion} 
	
	In this work we proposed the JGDLA as a method for approximate inference on Markov population models. We showed that the JGDLA produces a joint Gaussian distribution that relies solely on the solution to a system of ODEs. We showed that, unlike the Euler-Maruyama scheme, the JGDLA does not require stochastic infill to perform inference or predict at unobserved times. We also illustrated the connection between SDE and ODE modeling by demonstrating how to perform inference on Markov population models directly from a deterministic system in our simulation study and data analysis.  
	
	There are other frameworks for approximating the joint likelihood of the data, such as particle filtering \citep{doucet2009tutorial} and iterated filtering \citep{ionides2015inference}. These methods are commonly used when the Markov population process is assumed to be unobserved. Filtering methods rely on sequentially drawing samples from the filtering and predictive distributions of the latent states. This task requires recursively defining conditional densities from a Gaussian distribution. We fit a hidden Markov model in the COVID-19 data analysis of Section \ref{section::Covid Data Example} in which the population process was unobserved. The JGDLA does not require filtering methods to approximate the likelihood. We instead performed a Monte Carlo approximation of the likelihood using samples from the JGLDA's full joint distribution for the latent states.

	Inference for JGDLA models is not restricted to Bayesian approaches. In Section \ref{section::Simulation Study}, we considered a directly observed population process. This allowed for maximum likelihood estimates to be obtained by optimizing the JGDLA joint Gaussian likelihood. In Section \ref{section::Covid Data Example}, the JGDLA produces a joint distribution for the latent states at the time points at which seropositive data counts were observed. Maximum likelihood estimates could have been obtained by optimization, however we elected to use MCMC and found the results to be robust to initial conditions.  
	
	In summary, we have constructed the JGDLA as a new method for performing approximate inference on Markov population models. We showed that likelihood evaluations of the JGDLA depend solely on the solution of a system of ODEs. In turn, the JGDLA does not require stochastic infill to predict or perform inference. We saw the JGDLA approximation is built directly from a system of ODEs, connecting ODE and SDE modeling. We also observed that the JGDLA outperformed the Euler-Maruyama approximation in the SIR simulation study considered in Section \ref{section::Simulation Study}. In conclusion, we suggest the use of the JGDLA as a new framework for performing inference on Markov models.

	\bibliographystyle{Chicago}
	\bibliography{A_New_Framework_for_Inference_on_Markov_Population_Models}
	
	\begin{appendices}
		\section{Appendix}
		
		\subsection{FCLT for Poisson Processes} \label{appendix::FCLT for Poisson Processes} 
		
		We state the FCLT for Poisson processes used in Section \ref{section:Diffusion Approximaitons} to perform a Gaussian approximation of the Markov population model in \eqref{eqn::X PP sum}. The FCLT states that for a Poisson process with rate $\lambda t$, denoted $Y\left(\lambda t \right)$, as  $n \rightarrow \infty$ we have 
		\bea 
			\sqrt{n}\left( \frac{1}{n} Y(n \lambda t) - \lambda t \right) \Rightarrow B(\lambda t), \label{appendix::eqn::FCLT}
		\eea   
		where $B(t)$ is a standard Brownian motion, and $``\Rightarrow"$ denotes convergence in distribution \citep{kurtz1978strong, van1992stochastic}. We use \eqref{appendix::eqn::FCLT} to perform a Gaussian approximation of a Poisson process
		\bea 
			\frac{1}{n} Y(n\lambda t) \approx \lambda t + \frac{1}{\sqrt{n}} B(\lambda t), \label{appendix::eqn::PP Gaussian Approx} 
		\eea  
		for sufficiently large $n$.	
		
		\subsection{Solving for V(t)} \label{appendix::Solving for V(t)} 
		We seek solutions to \eqref{eqn::dV Linear Sigma} of the form 
		\bea 
			\bs{V}(t) = U(t)\bs{Y}(t), & & \bs{V}(0) = \bs{0}\nt,
		\eea 
		such that 
		\bea 
			dU(t) = \partial \bs{F}\left(\bs{X}^{\dagger}(t)\right)U(t), & & U(0) = \mathbb{I}_{d \times d}, \label{appendix::eqn::Matrix dUdt}
		\eea 
		and 
		\bea 
			d\bs{Y}(t) = K(t)d\bs{B}(t), & & \bs{Y}(0) = \bs{0}, \label{appendix::eqn::dY with K} 
		\eea 
		where $K(t)$ is $d$ by $n$ matrix that must be solved for. Using \eqref{appendix::eqn::Matrix dUdt} and \eqref{appendix::eqn::dY with K}, we have
		\bea 
			d\bs{V}(t)  &=& dU(t)\bs{Y}(t) + U(t)d\bs{Y}(t) \nt \\
						&=& \partial \bs{F}\left(\bs{X}^{\dagger}(t)\right)U(t)\bs{Y}(t)dt + U(t)K(t)d\bs{B}(t) \nt \\
						&=& \partial \bs{F}\left(\bs{X}^{\dagger}(t)\right)\bs{V}(t)dt + U(t)K(t)d\bs{B}(t). \label{appendix::eqn::dV(t)}
		\eea
		We ensure \eqref{appendix::eqn::dV(t)} agrees with \eqref{eqn::dV Linear Sigma} by taking
		\bea 
			U(t)K(t) = Q\left( \bs{X}^{\dagger}(t) \right) \Longrightarrow K(t) = U^{-1}(t)Q\left( \bs{X}^{\dagger}(t) \right).
		\eea 	
		The solution to the SDE in \eqref{eqn::dV Linear Sigma} is obtained by sequentially solving \eqref{appendix::eqn::Matrix dUdt} and 
		\bea 
			\bs{Y}(t) = \int_{0}^{t} U^{-1}(s)Q\left( \bs{X}^{\dagger}(s) \right) d\bs{B}(s). \nt 
		\eea

		\subsection{SIR JGDLA} \label{appendix::JGDLA for SIR}
		We denote the observed time points $\bs{X}^{obs}_N = \left(\bs{X}_N^{obs}(5), \bs{X}_N^{obs}(10), ..., \bs{X}_N^{obs}(30) \right)\p$. The deterministic SIR model was given in Section \ref{section::Simulation Study} 
		\bea 
			\frac{d}{dt} S^{\dagger}(t) &=& -\beta S^{\dagger}(t)I^{\dagger}(t), \label{appendix::eqn::S deterministic} \\
			\frac{d}{dt} I^{\dagger}(t) &=& \beta  S^{\dagger}(t)I^{\dagger}(t) - \gamma I^{\dagger}(t). \label{appendix::eqn::I deterministic}
		\eea  
		We first solve (\ref{appendix::eqn::S deterministic}--\ref{appendix::eqn::I deterministic}) for $\bs{X}^{\dagger}(t)= \left( S^{\dagger}(t), I^{\dagger}(t) \right)\p$ using initial conditions $\bs{X}_N(0)  = (0.95,0.05)\p$. We used the built-in numerical solver \textit{ode} in the \textit{deSolve} package of \textit{R} with a step size of 0.1, solver setting \textit{lsoda}, and initial guess $\theta_0 = (\beta_0,\gamma_0)$. We note that the results were robust to the choice of $\theta_0$. 
		
		We solve 
		\bea 
			dU(t) = \partial \bs{F}\left(\bs{X}^{\dagger}(t) \right) U(t), \hspace{0.1 in } U(0) = \mathbb{I}_{2 \times 2}, \label{appendix::eqn::dU}
		\eea 
		where
		\begin{equation}
			\partial \bs{F}\left(\bs{X}^{\dagger}(t) \right) = 
			\begin{bmatrix}
				-\beta I^{\dagger}(t) & -\beta S^{\dagger}(t) \\
				\beta I^{\dagger}(t) & \beta S^{\dagger}(t) - \gamma 
			\end{bmatrix}. \nt 
		\end{equation} 
		We solve \eqref{appendix::eqn::dU} numerically again using \textit{ode} with a time step of 0.1 and solver setting \textit{vode}. We interpolate $\bs{a}_i(s) = U^{-1}(s)\bs{R}_i$ across the time domain (0,30) and construct $\Sigma_Y$ by numerically solving 
		\bea 
			Var\left(Y_1(t)\right) &=& \int_{0}^{t} \left( \beta a^2_{11}(s)I^{\dagger}(s)S^{\dagger}(s) + \gamma a^2_{21}(s)I^{\dagger}(s) \right)ds, \label{appendix::eqn:cov y1} \\ 
			Var\left(Y_2(t)\right) &=& \int_{0}^{t} \left( \beta a^2_{12}(s)I^{\dagger}(s)S^{\dagger}(s) + \gamma a^2_{22}(s)I^{\dagger}(s) \right)ds, \label{appendix::eqn:cov y2} \\ 
			Cov\left(Y_1(t), Y_2(t)\right) &=& \int_{0}^{t} \left( \beta a^2_{11}(s)a_{12}(s)I^{\dagger}(s)S^{\dagger}(s) + \gamma a_{21}(s)a_{22}(s)I^{\dagger}(s) \right)ds, \label{appendix::eqn:cov y1 y2}
		\eea   
		at observed time points $t = 5,10,15,20,25,30$. Next, we construct $\Sigma^{\dagger}_{\bs{\theta}}$ in \eqref{eqn::Sigma Dagger} and evaluate the JGDLA likelihood
		\bea 
			\pi{\left( \bs{X}^{obs}_N | \bs{X}^{\dagger}, \bs{\theta} \right)} \sim N\left(\bs{X}^{\dagger}, \Sigma^{\dagger}_{\bs{\theta}} \right), 
		\eea  
		where $\bs{X}^{\dagger} = \left(\bs{X}^{\dagger}(5), \bs{X}^{\dagger}(10), ..., \bs{X}^{\dagger}(30) \right)\p$. We use \textit{optim} in \textit{R} to find maximum likelihood estimates by performing the process above iteratively for differing values of $\bs{\theta}$. 
		
		\subsection{SEIR JGDLA} \label{appendix::JGDLA for SEIR}
		
		We denote the fourteen times at which COVID tests $n_t$ were administrated yielding $y_t$ positive tests $T^{obs} = \{1,2,3,..,6,8,9,11,12,...,16\}$. We denote the solution to the system of ODEs in Section \ref{section::Covid Data Example} at time points $T^{obs}$, $\bs{X}^{\dagger} = \left(\bs{X}^{\dagger}(1),...,\bs{X}^{\dagger}(6),\bs{X}^{\dagger}(8),\bs{X}^{\dagger}(9),\bs{X}^{\dagger}(11),...\bs{X}^{\dagger}(16)\right)\p$. We detail the construction of the JGDLA for the latent proportions $\bs{X}_N = \left(\bs{X}_N(1),...,\bs{X}_N(6),\bs{X}_N(8),\bs{X}_N(9),\bs{X}_N(11),...\bs{X}_N(16)\right)\p$.
		
		We first solve for $\bs{X}^{\dagger}$ conditioned on $\bs{X}_N(0) = \left( S_N(0), E_N(0), I_N(0) \right)\p$ and $\bs{\theta} = \left(\beta,\alpha,\gamma\right)$ using the built-in numerical solver \textit{ode} in the \textit{deSolve} package of \textit{R} with a step size of 0.1 and solver setting \textit{vode}. Next we define  
		\begin{equation} 
				\partial \bs{F}\left(\bs{X}^{\dagger}(t)\right) = \begin{bmatrix}
							-\left( \beta I^{\dagger}(t) + \mu_S(t) \right) & 0 & -\beta S^{\dagger}(t) \\
							\beta I^{\dagger}(t) & -\alpha & \beta S^{\dagger}(t) \\
							0 & \alpha & -\gamma 
			\end{bmatrix}.
		\end{equation} 
		and solve for $U(t)$ in \eqref{eqn::dU} using \textit{ode} with a step size of 0.1 and solver setting \textit{vode}. We interpolate $\bs{a}_i(s) = U^{-1}\bs{R}_i$ on (0,16] and use the result to numerically solve each in integral of \eqref{eqn::Cov Y} using \textit{ode} with a time step of 0.1 and numerical solver setting \textit{vode} for time points $t \in T^{obs}$. We then form $\Sigma^{\dagger}_{\bs{\theta}}$ in \eqref{eqn::Sigma Dagger} to obtain our JGDLA density $\pi{\left( \bs{X}_N | \bs{X}^{\dagger},\bs{\theta} \right)}$. 
		
		We fit the stochastic SEIR model of Section \ref{section::Covid Data Example} via MCMC. We note that for each iteration of MCMC, we propose $\bs{\theta}^{*}$ and $\bs{X}^{*}_N(0)$ all-at-once. Note that we propose $S^{*}(0)>0$ and $I^{*}(0)>0$ under the constraints $S^{*}_N(0) + I^{*}_N(0)  < 1$, $R_N(0) = 0$, and $E^{*}_N(0) = 1-S^{*}_N(0) - I^{*}_N(0)$. For each proposed value of $\bs{\theta}^{*}$ and $\bs{X}^{*}_N(0)$, the JGDLA must be constructed to obtain $\pi{\left( \bs{X}_N | \left( \bs{X}^{\dagger}\right)^{*},\bs{\theta}^{*} \right)}$, where $\left( \bs{X}^{\dagger}\right)^{*}$ is the ODE solution resulting from initial conditions $\bs{X}^{*}_N(0)$ and $\bs{\theta}^{*}$. We then perform a Monte Carlo approximation for $\hat{\pi}{\left( \bs{y} | \bs{\theta}^{*} \right) }$ by drawing 1,000 samples from $\pi{\left( \bs{X}_N | \left( \bs{X}^{\dagger}\right)^{*},\bs{\theta}^{*} \right)}$ and following the details of Section \ref{section::Covid Data Example}. A Metropolis-Hasting accept/reject step is then performed with posterior conditional distribution $\pi{\left( \bs{\theta},\bs{X}_N(0) | \bs{y} \right)} \propto \hat{\pi}{\left(\bs{y} | \bs{\theta} \right)}\pi{\left( \bs{X}_N(0) \right)}\pi{\left(\bs{\theta}\right)}$.

	\end{appendices}
	
\end{document}